\newcommand{\beq}{\begin{equation}}
\newcommand{\eeq}{\end{equation}}
\newcommand{\beqa}{\begin{eqnarray}}
\newcommand{\eeqa}{\end{eqnarray}}

\renewcommand{\lambda}{\ell}

\newcommand{\bk}{{\bf k}}
\newcommand{\br}{{\bf r}}

\documentstyle[aps,prl,twocolumn,epsf]{revtex}

\begin{document}
\pagestyle{arabic}

\twocolumn[\hsize\textwidth\columnwidth\hsize\csname @twocolumnfalse\endcsname

\hfill{LA-UR-98-2174}

\title{Marginal stability of d-wave superconductor: spontaneous P and T 
violation 
in the presence of magnetic impurities \footnotemark[1]}

\author{ A.V.~Balatsky $^{1}$ and R.~Movshovich $^{2}$}
\address{T-Div and MST-Div, Los Alamos National Laboratory, Los Alamos, 
New Mexico 87545, USA}
  
\date{\today}
\maketitle

\begin{abstract}
We argue that the $d_{x^2-y^2}$-wave superconductor is marginally stable in the 
presence of external perturbations.  Subjected to the external perturbations by  
magnetic impurities, it develops a secondary component of the gap, {\em complex} 
$d_{xy}$,  to maximize the coupling to impurities and lower the total energy. 
The 
secondary $d_{xy}$ component exists at high temperatures and produces the full 
gap 
$\sim 20 K$ in the single particle spectrum around each impurity, apart from 
impurity induced broadening. At low temperatures the phase ordering transition 
into global $ d_{x^2-y^2} + i d_{xy}$ state occurs. 
\
 
 
\end{abstract}

\

]


The point of this note is to emphasize the recently recognized new
aspect of the high temperature superconductors :{\em  a marginal stability of 
the 
$d_{x^2-y^2}$-wave
superconductor towards secondary ordering in the presence of the
symmetry perturbing field}. Namely: in the presence of the perturbing
field the $d_{x^2-y^2}$- wave superconductor generates the secondary
superconducting component of the order parameter, likely to be  $id_{xy}$
 in our case,   to maximize the
coupling to this field and hence to lower the total energy. 
\footnotetext[1]{Work
 done in 
collaboration with  M.A. Hubbard (UIUC), M.B. Salamon (UIUC), R. Yoshizaki 
(Univ. of Tsukuba), J. Sarrao (LANL)
 and M. Jaime (LANL)}

 This instability can occur in many different ways.  Recently the 
surface-scattering induced s-wave component in high-T$_c$ materials has been 
observed  \cite{Greene1} and the model explaining the effect was proposed 
\cite{Fogel1}. The existence of the secondary gap in the external magnetic  
field 
was suggested  to explain  the anomalies in thermal transport in Bi2212 
\cite{Ong1,Laugh1}. In both of these cases the superconductor was subjected to 
the 
perturbing fields: the surface scattering or the external magnetic field. The 
above examples can be thought of as  a specific realizations  of  the general 
phenomena of marginal stability of  $d_{x^2-y^2}$-wave
superconductor.

Specifically we 
investigated
the role of magnetic and nonmagnetic impurities on the $d_{x^2-y^2}$-wave 
superconductors. We find  that in the vicinity of each magnetic impurity, in the 
presence of the spin-orbit coupling, there is a patch of local complex $id_{xy}$ 
gap generated from impurity scattering. This is the first example, to the best 
of 
our knowledge, when impurity scattering produces the coherent component, i.e.  
secondary $id_{xy}$ gap, as shown in Fig.1. We suggest  that the secondary phase 
transition $d_{x^2-y^2} \rightarrow  d_{x^2-y^2} + i d_{xy}$ occurs 
spontaneously 
at lower temperatures with simultaneous impurity spin ordering. Below we present 
the summary of the results using mostly qualitative description. For more 
technical approach reader is advised to look at the original papers 
\cite{avb1,roman1}.

 {\em a) Single magnetic impurity}. The essence of the argument is to consider 
the 
single magnetic impurity
with large spin S in the $d_{x^2-y^2}$-wave superconductor. Locally the
time reversal     (T) and parity (P) symmetries are violated as the
direction of the spin is fixed. Consequently, in the presence of the
symmetry perturbing field it is favorable for superconducting state to
generate the secondary component, i.e. the  $id_{xy}$, so that the
superconducting condensate couples to the impurity spin and lowers the
total energy. 

\begin{figure}
\epsfxsize=2.2in
\centerline{\epsfbox{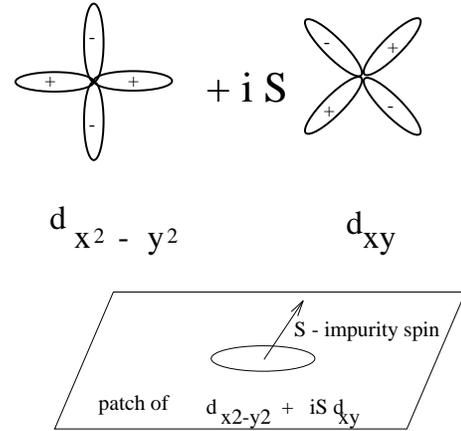}}
\caption[]{ The P and T violating condensate in the presence of magnetic 
impurity 
is shown.  The phase of the 
induced
$d_{xy}$ component is determined by the $S_z$,
the impurity spin. At high temperatures 
the phase of induced component is disordered due to spin flips.  
At low temperatures  the Josephson tunneling  locks the  phase between patches, 
leading to the global $d_{x^2-y^2}+id_{xy}$ state. }
\end{figure}

Consider scattering of a $d_{x^2-y^2}$ pair off the single impurity site. 
Interaction Hamiltonian is $H_{int} =  g L_z S^z$, $L_z = i \hbar 
\partial_{\theta}$  is the  angular momentum operator, $\theta$ is the angle 
on 
the cylindrical 2D Fermi surface, $S_z$ is the out of  plane component of the 
impurity spin and $g$ is the spin orbit coupling constant. There is   a finite 
scattering amplitude  $\langle x^2-y^2|H_{int}|xy\rangle$ in the vicinity of 
impurity:
\beqa
\langle \Delta_0 \cos 2\theta|i g S^z\hbar \partial_{\theta}|\Delta_1 
\sin2\theta 
\rangle \sim i  S^z\Delta^*_0\Delta_1
\label{Jos1}
\eeqa
where $x^2-y^2 \sim \cos 2\theta$ and $xy \sim \sin 2\theta$ order parameter 
amplitudes are $\Delta_0$ and $\Delta_1$
respectively. This scattering amplitude  $\langle x^2-y^2|H_{int}|xy\rangle$ 
does 
imply the existence of the finite $d_{x^2-y^2} + id_{xy}$ gap near each Ni site.  
The global second phase grows out of these patches at lower temperatures.

The precursors of the ordered phase, i.e. a {\em finite} quasiparticle gap near 
each impurity site should be seen even at  temperatures above the second 
transition into  $d_{x^2-y^2}+id_{xy}$ state. 

 In the presence of the 
single impurity scattering potential: $F_{\omega_n}(\bk,\bk') = 
F^0_{\omega_n}(\bk)\delta(\bk-\bk') + F^1_{\omega_n}(\bk,\bk')$, where
$F^0 = {\Delta_0 \cos 2\theta\over{\omega_n^2 + \xi_{\bk} + \Delta_0^2 \cos^2 2 
\theta}}, G^0 = -{i\omega_n + \xi_{\bk}\over{\omega_n^2 + \xi_{\bk} + \Delta_0^2 
\cos^2 2 \theta}}$ are the pure system propagators, $F^1_{\omega_n}(\bk,\bk')$ 
is the correction due to impurity scattering,  $\bk = (k, \theta)$ are the 
magnitude and angle of the momentum $\bk$ on the  cylindrical Fermi surface, 
$\omega_n$  is Matsubara frequency and  $\xi_{\bk} = \epsilon_{\bk}-\mu$ is the 
quasiparticle energy, counted form the Fermi surface.
To linear order in small 
$gN_0$ ($N_0$ is the density of states at the Fermi surface),  one finds  
\cite{avb1} :
\beqa
F^1_{\omega_n}(\bk,\bk') =  -i2\pi g S_z G^0_{\omega_n}(\bk) 
F^0_{\omega_n}(\bk') {[\bk \times \bk']_z\over{|\bk-\bk'|}}
\label{F1}
\eeqa
Where $ F^1_{\omega_n}(\bk,\bk') $ is the function of incoming and outgoing 
momenta because of broken translational symmetry. The first nontrivial 
correction 
to the homogeneous solution, after integration over $\bk'$ and $\xi_{\bk}$, is 
the 
$xy$ component:
\beqa
F^1_{\omega_n}(\theta)=\int N_0 d\xi_{\bk} F^1_{\omega_n}(\bk) \propto i   
 (N_0 gS_z)(N_0 \Delta_0)\sin 2 \theta
\label{F1int}
\eeqa
The finite induced $xy$ component of the order parameter also leads to the $xy$ 
gap:
\beqa
\Delta_{1}(\bk,\bk'') = T\sum_{n,\bk'} 
V_{xy}(\bk,\bk')F^1_{\omega_n}(\bk',\bk'') 
\nonumber\\
\Delta_{1} \propto 2\pi g (N_0\Delta_0)(N_0 V_{xy}) \sim 20 K
\label{D1}
\eeqa
 Where  $V_{xy}(\bk,\bk')N_0$ is the arbitrary sign interaction in the  $xy $ 
channel, assumed to be of $V_{xy}N_0 \sim 0.1$ strength .

The finite minimal gap on the Fermi surface near impurity  site 
is determined by: $ Gap = \sqrt{|\Delta_0(\theta)|^2 + |\Delta_1(\theta)|^2} 
\sim 
20 K$. Experimental prediction following from this picture is that the 
pseudogapped particle spectrum with minimal gap on the order of 20 K should be 
seen in scanning tunneling microscope 
experiments near each impurity site.

 The usual impurity induced   broadening of the states will be present as well. 
At 
low concentrations the broadening, being function of impurity concentration will 
be small compared to the induced $d_{xy}$ gap.  See also Fig.3.

{\em b) Finite impurity concentration}.

Recently Movshovich et.al. \cite{roman1} measured thermal transport in high 
temperature superconductor with magnetic impurity  (Bi2212 with Ni). The 
surprising outcome of these 
experiments is the observed sharp reduction in thermal conductivity at $0.2 K$ 
in 
the samples with $1-2\% \%$ Ni impurity concentration (see Fig. 2). The observed 
feature in thermal conductivity is consistent with the second superconducting 
transition into a $d_{x^2-y^2} + i d_{xy}$ state as described.  The secondary 
phase in many respects resembles the superfluid $^3He$: this is a chiral state 
that violates P and T. The superconducting  condensate  has a nonzero orbital 
moment $L$.

\begin{figure}
\epsfxsize=2.2in
\centerline{\epsfbox{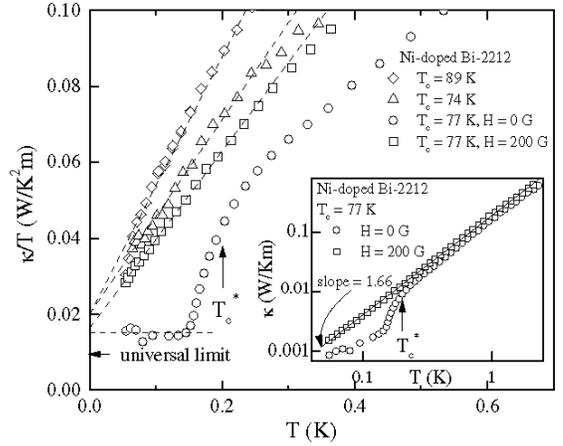}}
\caption[]{The thermal conductivity of the Ni-doped Bi2212 is shown. The sharp 
reduction of thermal conductivity occurs at $T^*_C = 0.2 K$. The inset shows the 
effect of the applied magnetic field that suppresses the feature. The data are 
consistent with the secondary superconducting phase, such as $d_{x^2-y^2} + i 
d_{xy}$, developing
at 
$T^*_C$ and with the full gap opening up.  No effect has been observed 
in
 the nonmagnetic impurity (Zn) doped samples available to us. }
\end{figure}

 The free energy admits the linear coupling between the $d_{x^2-y^2}$
and $d_{xy}$ channels, see Eq.(\ref{Jos1}):
\beqa F_{int} = i \Delta_0\Delta^*_1 S_z + h.c.
\label{int}
 \eeqa
 which can be thought of as a spin-assisted Josephson coupling between
 orthogonal $x^2-y^2$ and $xy$ channels. Since all other relevant terms are 
quadratic and higher powers in $\Delta_1$ and $S_z$, this linear coupling is 
driving the transition into $d_{x^2-y^2}+id_{xy}$ state.

 Impurities, in addition to the $d_{xy}$ component  produce the finite lifetime 
for quasiparticles. Standard arguments of Abrikosov-Gorkov theory imply that the 
transition temperature into $d_{x^2-y^2}$ is suppressed. Moreover, the same 
impurity scattering will suppress the secondary transition into 
$d_{x^2-y^2}+id_{xy}$ state. We expect there is a {\em finite} impurity  
concentration window where induced phase can exist, see Fig.3.

\begin{figure}
\epsfxsize=2.2in
\centerline{\epsfbox{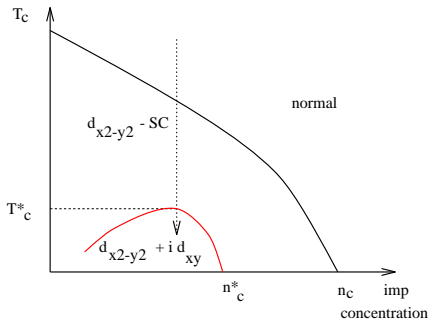}}
\caption[]{ The suggested phase diagram for the normal, $d_{x^2-y^2}$ and 
$d_{x^2-y^2} + i d_{xy}$ phases  as a function of impurity concentration is 
shown. 
At low impurity concentration the patches of $d_{x^2-y^2} + i d_{xy}$
presumably  can order, although at very low temperatures. The low impurity 
concentration cut off will be determined by quasiparticle scattering The 
termination 
point at $n^*_c$, when the impurity scattering will suppress $T^*_c$ to zero, is 
expected. For $n \leq n^*_c$ the sequence of the transitions upon temperature 
lowering is shown by the dotted line. Note that the temperatures $T_c$ and 
$T^*_c$ 
are drawn out of scale. 
 }
\end{figure}

The idea of marginal stability of high temperature superconductors and of the 
secondary superconducting phase might have a broader application for other 
unconventional superconductors such as heavy fermion compounds. It implies that 
the superconducting phase diagram in many of these compounds might be richer 
than 
we previously thought.

Observation of such a state would 
represent a significant new development in the field of high temperature 
superconductivity.

The useful discussions with E. Abrahams,  L. Greene,   R. Laughlin, D.H. Lee, M. 
Salkola and  J. Sauls are gratefully acknowledged. This work was supported by US 
DoE.


\begin{references}




\item[]{$^1$ -- avb@lanl.gov, $^2$ -- roman@lanl.gov}






\bibitem{Greene1}  M. Covington, et.al., Phys. Rev. Lett., {\bf 79}, 277, 
(1997).

\bibitem{Fogel1}  M. Fogelstrom et.al., Phys. Rev. Lett., {\bf 79}, 281, (1997). 

\bibitem{Ong1}  K. Krishana, et.al., Science, {\bf 277}, 83, (1997). See also H. 
Aubin, K. Behnia, S. Ooi, T. Tamegai; K. Krishana,
         N. P. Ong, Q. Li, G. Gu, N. Koshizuka
           Science 1998 April 3; 280 (5360):11 (in Technical
         Comments).
         

 \bibitem{Laugh1} R.B. Laughlin, Preprint,  cond-mat/9709004. 


\bibitem{avb1}  A. V. Balatsky, Phys. Rev. Lett, {\bf 80},1972 (1998). Also 
cond-mat/9710323.


\bibitem{roman1} R. Movshovich et.al., Phys. Rev. Lett,  {\bf 80}, 1968 (1998). 
Also cond-mat/9709061.



\end{references}
\end{document}